\documentclass{PoS}

\usepackage{subcaption} 
\usepackage{multirow} 
\usepackage{siunitx} 

\newcommand*\Fermis{\textit{Fermi}-LAT }

\newcommand{\similar}{\ensuremath{\sim}}
\newcommand{\pfrac}{$e^{+} / (e^{+} + e^{-})$}
\newcommand{\aprat}{$\overline{p}/p$}
\newcommand{\GR}{$\gamma$-ray}

\title{Observing The Cosmic Ray Moon Shadow With VERITAS}

\ShortTitle{Observing The Cosmic Ray Moon Shadow With VERITAS}

\author{\speaker{Ralph Bird}, for the VERITAS Collaboration\thanks{http://veritas.sao.arizona.edu}\\
        University College Dublin\\
        E-mail: \email{ralph.bird.1@gmail.com}}

\abstract{The Earth is subjected to a uniform flux of very-high-energy (VHE, E~$>$~\SI{100}{GeV}) cosmic rays unless they are obscured by an object, such as the Moon, in which case a deficit or shadow is created. 
Since cosmic rays are charged this deficit is deflected by the Earth's magnetic field, enabling the rigidity of the obstructed cosmic rays to be determined. 
Measurement of the relative deficits of different species enables the positron fraction and the antiproton ratio to be measured. 
The April 15, 2014 lunar eclipse was visible with the VERITAS Cherenkov telescopes, which allowed (with special UV bandpass filters) 74 minutes of direct observations of the Moon and the associated deficit in the cosmic-ray flux. 
The results of this observation are presented. 
In addition VERITAS has been conducting a series of observations by pointing close to a partially illuminated Moon, with a reduced photomultiplier tube high voltage and UV bandpass filters. 
We present the technique developed for these observations and their current status.}

\FullConference{The 34th International Cosmic Ray Conference,\\
		30 July- 6 August, 2015\\
		The Hague, The Netherlands}

\begin{document}
\section{Inroduction}
Since their discovery over 100 years ago, cosmic rays have been the subject of intense research, yet their origin remains a mystery.  
Key evidence to help solve this problem will come from the energy spectrum of cosmic rays and the relative proportions of the different cosmic ray types, in particular the ratio of antimatter to matter.
In this work we are looking at two important measures: the positron fraction (\pfrac) in the range \similar~0.5 to \similar~{1}{TeV} and the antiproton ratio (\aprat) in the \similar~1.5 to \similar~{3}{TeV} range.
This is done using the Earth-Moon ion spectrometer (EMIS), pioneered for use with IACTs by the ARTEMIS collaboration \cite{Pomar2001} and also being used by the MAGIC collaboration to measure the positron fraction \cite{Colin2009, Colin2011b}. 
Observations are subject to constraints on Moon elevation and illumination, thus observing time is limited to a few hours per year, this time includes observations during a Lunar eclipse on 2014-04-15.

\section{Positron Fraction \& Antiproton Ratio}
\begin{figure}[h]
\centering
\includegraphics[width=0.8\linewidth]{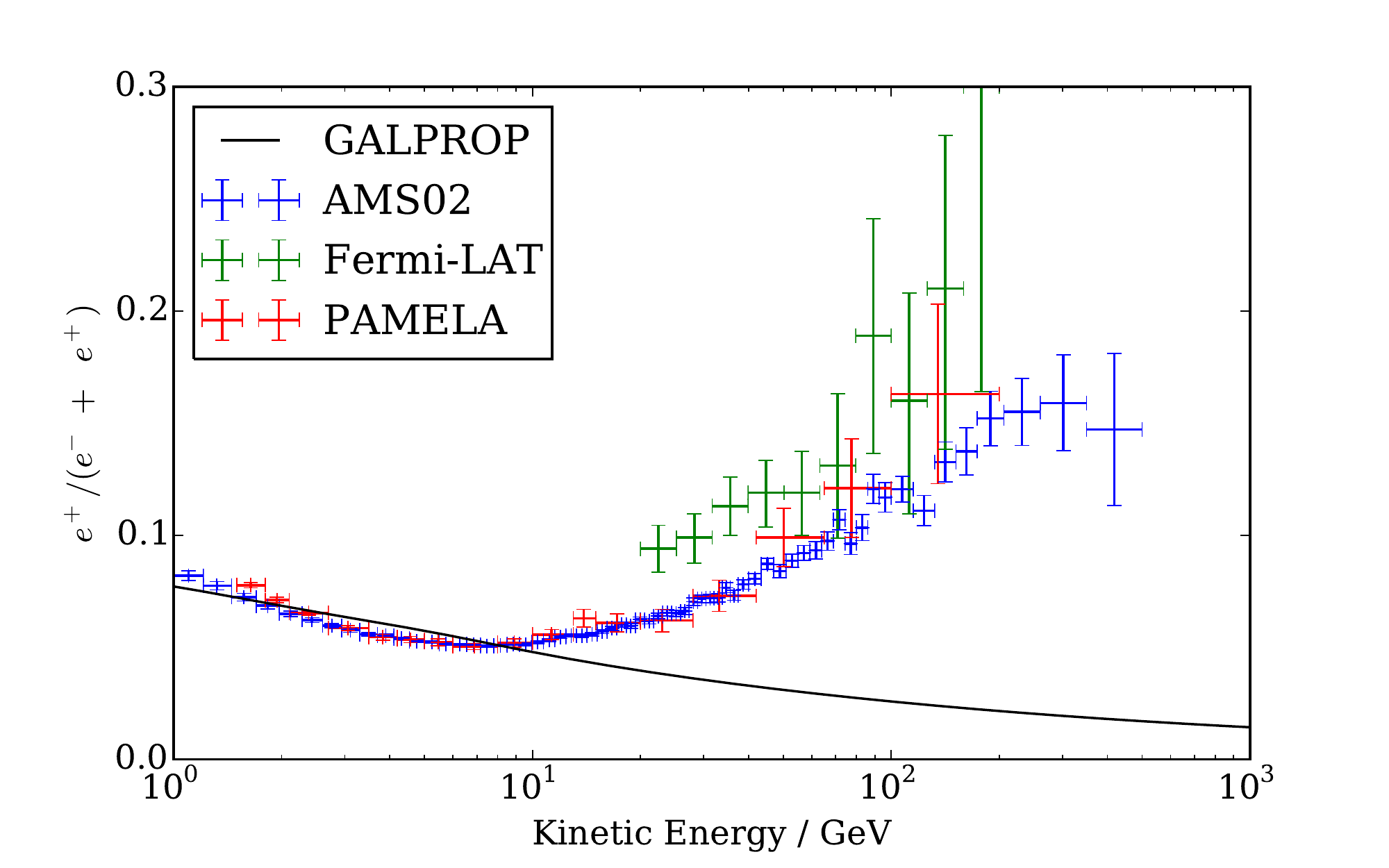}
\caption{Current measurements of the positron fraction from AMS-02 \cite{Aguilar2013}, PAMELA \cite{Adriani2009} and the \Fermis \cite{Ackermann2012}.  Overlaid is a typical GALPROP prediction \cite{GALPROP} of the positron fraction showing the divergence above \similar\SI{10}{GeV}.  Data is from \cite{Maurin2014}.}
\label{Fig:PosFrac}
\end{figure}

Since the first conclusive result of an increase in the positron fraction from the PAMELA collaboration \cite{Adriani2009} and especially with the improved statistics and energy range provided by AMS-02 \cite{Aguilar2013} ,the positron fraction has been of particular interest.
This rise was largely unexpected and deviates markedly from ``traditional'' models as shown in Figure \ref{Fig:PosFrac} where a typical GALPROP \cite{GALPROP} result is overlaid.
The key question has been what will happen at higher energies where a measure of the fraction can greatly constrain the existing models.

\begin{figure}[h]
\centering
\includegraphics[width=0.7\linewidth]{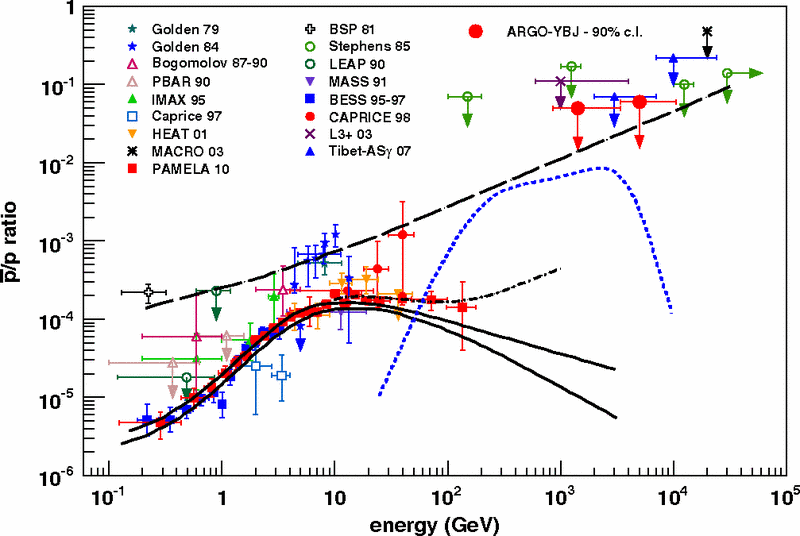}
\caption[Antiproton Ratio]{The antiproton ratio (from \cite{Bartoli2012}) showing the upper limits from a variety of experiments above \SI{1}{TeV}, the solid lines are predictions based upon pure secondary production \cite{Donato2009} and the dot-dash line includes contributions from additional sources of $\bar{p}$ at cosmic ray accelerators \cite{Blasi2009a}.}
\label{Fig:APRatFull}
\end{figure}

The antiproton ratio is well measured up to a few hundred GeV (with data from AMS-02 recently reducing the statistical uncertainty and increasing the energy range).
However, above \SI{1}{TeV} there are only upper limits, as can be seen in Figure \ref{Fig:APRatFull}.
At present, none of  these limits are constraining on typical models that fit the existing antiproton ratio measurements at lower energies.

\section{The Earth-Moon Ion Spectrometer}
\begin{figure}[h]
\centering
\includegraphics[width=0.4\linewidth]{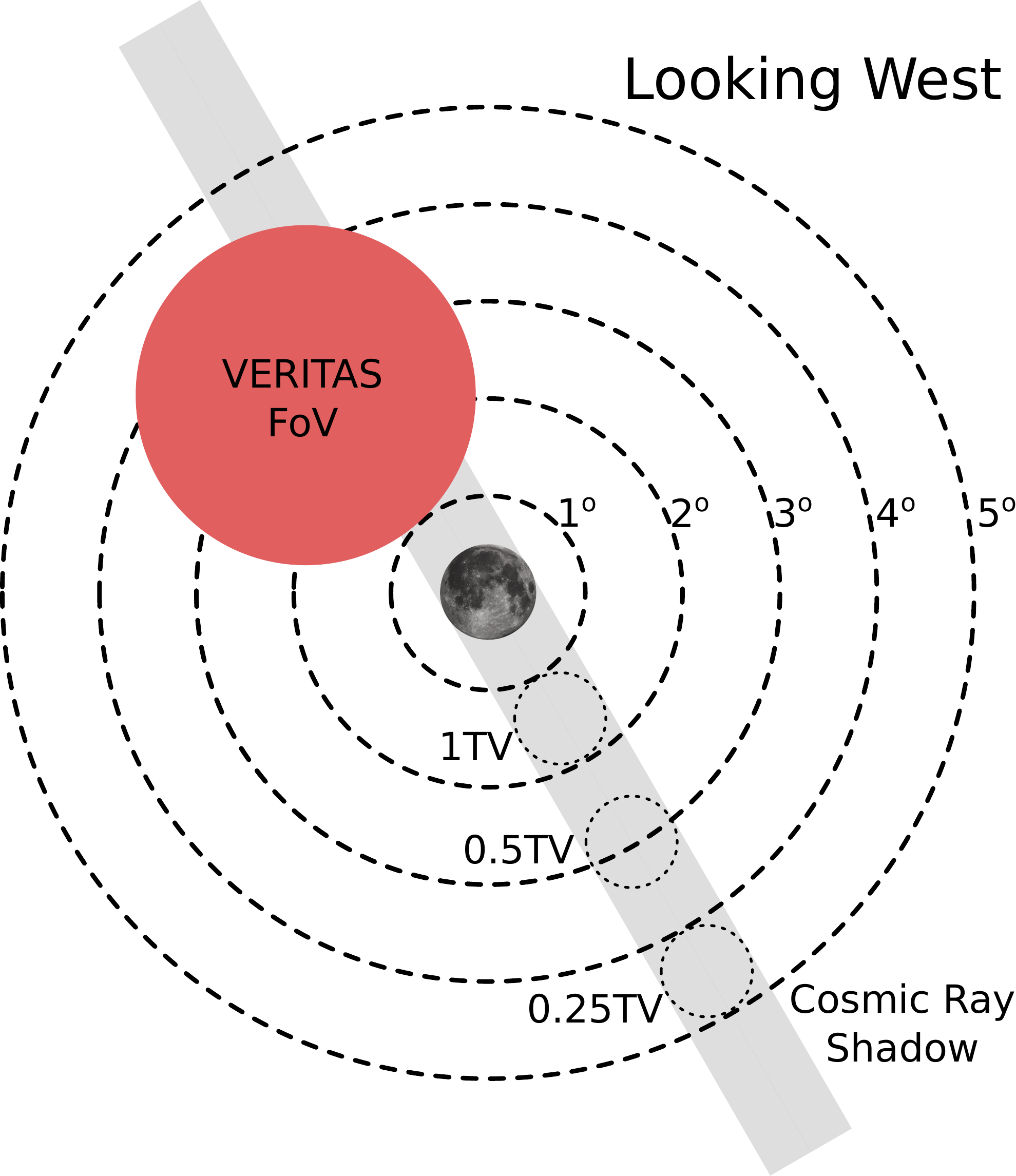}
\caption{A diagram showing the cosmic ray shadow the amount of deflection (shown for various rigidities as dotted circles) depends on the position in the sky.  For a falling Moon in the west, the positive shadow is at a lower elevation.}
\label{Fig:MSDiagram}
\end{figure}

The Earth-Moon ion spectrometer (EMIS) is a technique to determine the relative fluxes of cosmic rays of different rigidities ($R = p/q$ where $p$ is the particle momentum, $q$ is the charge).
The basis of the technique is that the cosmic ray flux is uniform across the sky.  
The only exception is if there is an object obstructing this flux, creating a deficit or shadow, such as the Moon.
Since cosmic rays are deflected by the Earth's magnetic field, with the amount of deflection depending upon the rigidity of the cosmic ray, this deficit will also be deflected.
When viewing this shadow from the ground (Figure~\ref{Fig:MSDiagram}) it creates a band that passes across the Moon, with one charge deflected one way, one the other and with lower rigidity particles deflected further than higher rigidity particles.
The ratio between the strength of the positive and negative shadows can be used to determine the ratio between positive and negative species.
By applying analysis cuts, hadronic and leptonic showers can be differentiated, allowing for separate ratios to be determined for the two.

\section{Feasibility of Observing with VERITAS}
\subsection{Bright Moonlight Observations}
In order to conduct these observations the telescopes must be pointed close to a partially illuminated Moon.  
The Moon is a significant light source that will increase the background light, reducing the signal-to-noise ratio, increasing the trigger rate (and thus the deadtime) and potentially  damaging the PMTs.  
To reduce its impact two techniques were employed, the high voltage supplied to the PMTs was reduced (RHV) and Schott glass UG-11 filters (UVF) were employed that only pass UV light, reducing the amount of light entering the PMTs.
These methods and their impact on the performance of the array are discussed in detail elsewhere in this conference \cite{Griffin2015}.
For the normal Moon shadow observations, a combined approach with both RHV and UVF is used.
During the 2014 lunar eclipse, due to the reduced lunar brightness we were able to observe using only UVF.

\subsection{Shadow Position}
In order to conduct the observations we need to know where the shadow is. 
The deflection of a particle is a function of the magnetic field, with the amount of deflection depending upon the rigidity of the particle.
In order to find this position we used the code TJ2010 \cite{TJ2010}.
The shadow axis is defined using the Moon and the position of the $\pm\SI{750}{GV}$ shadow.
Observations are conducted with the centre of the camera pointed $\pm\ang{0.5}$ perpendicular to this line. 
Since the shadow is the width of the Moon (\ang{0.5}) and with an angular resolution of \ang{0.1} for leptonic showers, this pointing means the shadow is fully contained in one half of the camera leaving the other half for background estimation.

\subsection{Crab Nebula Observations}
On 2013-03-19 the Crab Nebula was \similar\ang{3} from a 45\% illuminated Moon, this provided an opportunity to test the sensitivity of the array in the proposed observing conditions (UVF + RHV). 
Observations were conducted for all of the time the Crab Nebula was above \ang{40} elevation, with a total live time of \SI{107.5}{minutes}.
These data show that even with both RHV and UVF the VERITAS array is still sensitive to \GR s, though at a much reduced level of 6.82 $\sigma / \sqrt{hours}$ using the standard analysis cuts.
For comparison, normal observations have a sensitivity of around 30 $\sigma / \sqrt{hours}$ at these elevations.

\subsection{Observation Simulation}
\begin{figure}[h]
\centering
\includegraphics[width=0.8\linewidth]{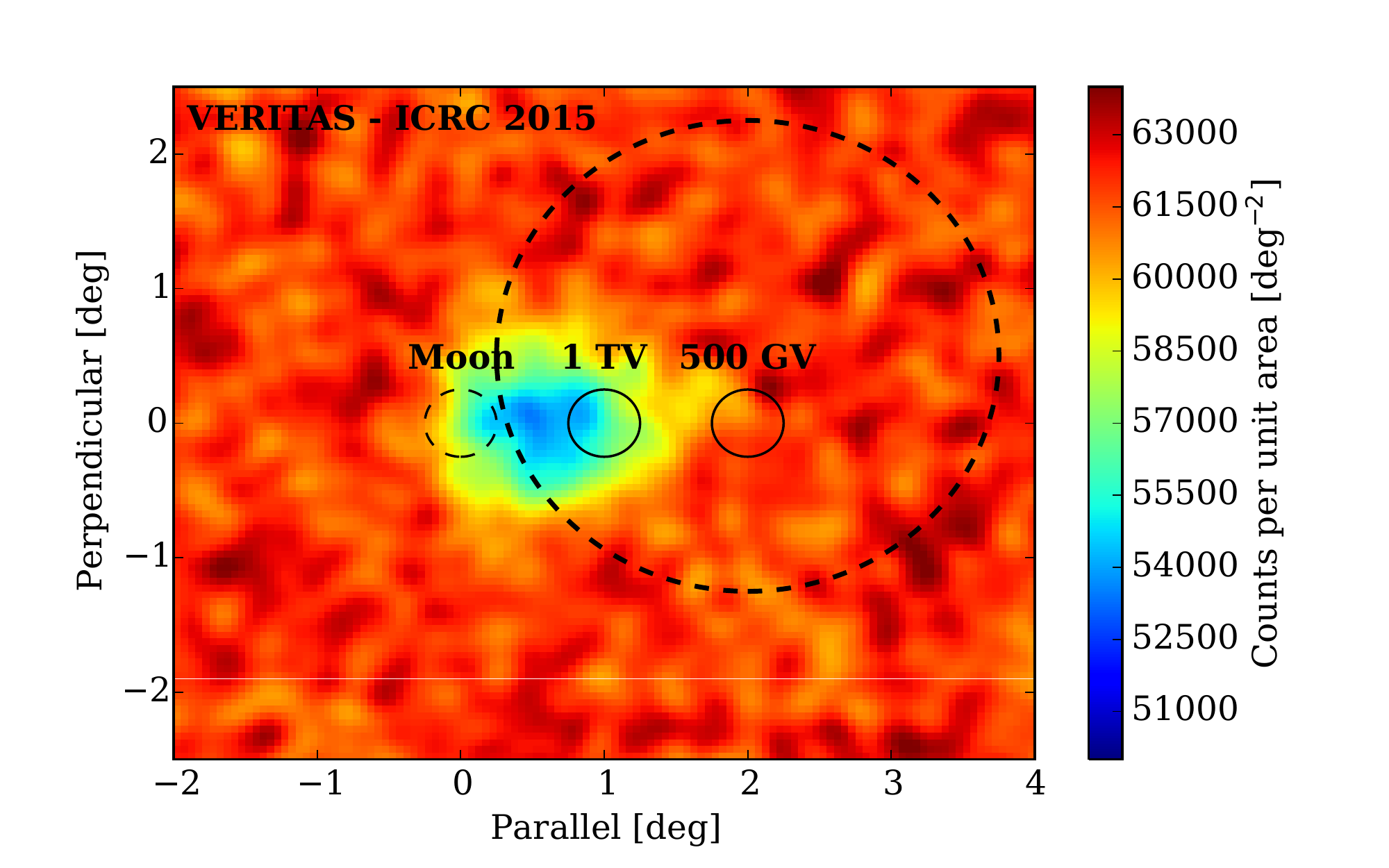}
\caption{Moon shadow simulation, the number of events within the FoV (large dashed circle) is approximately equivalent to 30~hours of observations. The shadow shape is influenced by the large angular resolution (\ang{0.4}), high energy threshold (\SI{1.5}{TeV}) and the source spectrum (a power law of index -2.7) which results in the shadow lying close to the Moon in a ``blob'' rather than an band.}
\label{Fig:HadSimSmooth}
\end{figure}

In order to test the observation's feasibility and to develop the analysis techniques, a Monte-Carlo model was generated assuming an energy threshold of \SI{500}{GeV} for leptons and \SI{1.5}{TeV} for hadrons.
This simulation (smoothed with a Gaussian kernel of \ang{0.1}) is shown in Figure \ref{Fig:HadSimSmooth}, the total number of counts falling within on field-of-view (FoV) is \similar$2.2\times10^6$.
This corresponds to 30 hours of observations at a reconstruction rate of \SI{10}{Hz}. 
The proton shadow is clearly visible cutting off sharply due to the power law spectrum just before \SI{1}{TV} shadow position, along with the fainter helium shadow which extends through the \SI{1}{TV} shadow position to the \SI{500}{GV} position due to its factor or two lower rigidity. 
The angular resolution (which is much larger at \ang{0.4} for hadronic showers) extends the shadow in all directions and will be a major constraint in this analysis since it will reduce the sensitivity by smearing out the signal and reducing the area available for estimating the background.

\begin{figure}[h]
\centering
\includegraphics[width=0.8\linewidth]{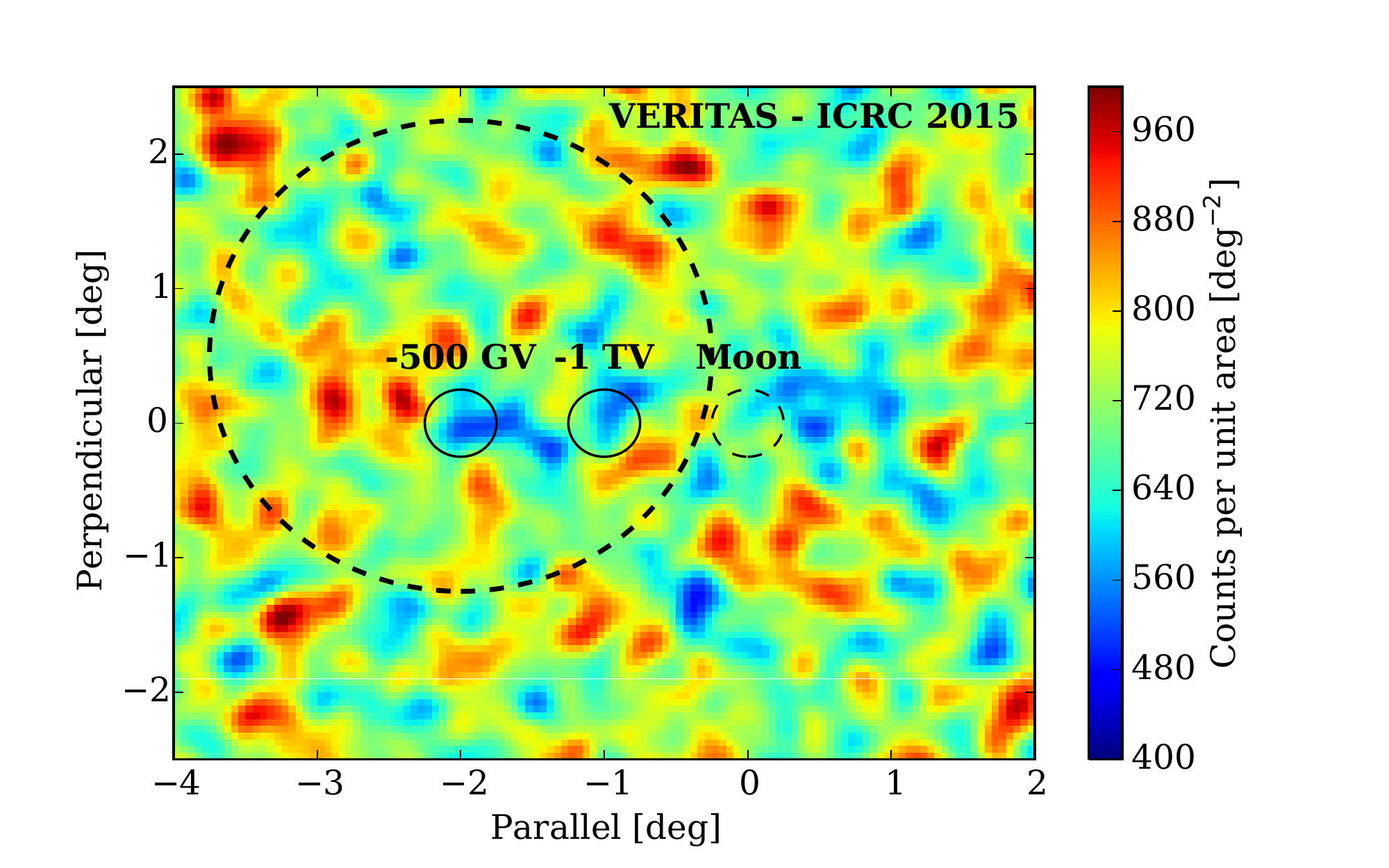}
\caption{A simulation of the expected skymap after gamma/hadron selection cuts are applied.  The number of events within the FoV is approximately equivalent to 30~hours of observations.}
\label{Fig:LeptSimSmooth}
\end{figure}

If selection cuts are applied, the survival fraction is 0.3 for leptons and 0.001 for hadrons, a simulation with the counts reduced to these fractions is shown in Figure \ref{Fig:LeptSimSmooth}.
The electron deficit just above the threshold energy is  visible at a rigidity of -500 to \SI{-700}{GV}, though a lot weaker than the proton deficit in the previous simulation.
With the lower statistics, statistical fluctuations produce comparable deficits elsewhere in the simulation, though the one associated with the electron shadow is present in all simulations, thus the 30 hours of observations can be considered a minimum.
The positron deficit is not visible and requires significantly more time to detect using this simple analysis.
However, neither simulation has made use of the reconstructed energy to aid in the analysis, the addition of this parameter will significantly improve the sensitivity.

\subsection{Feasibility Study Conclusions}
This study shows that it may be feasible to detect the shadow with VERITAS.
Taking tens of hours to detect either shadow (potentially hundreds for the positron shadow) it will require a multiyear investment as a maximum of $15-20$ hours of observing time can be expected in any one year (less with bad weather).
However, since an upper limit on the positron fraction in the 0.5 to \SI{1}{TeV} range could potentially provide useful constraints, a detection of the positron shadow may not be required.
Observations and support work are ongoing.

\section{Normal EMIS Observations}
Observations taken during the 2013-14 observing season totalled 326~minutes (out of a maximum possible of \similar900~minutes). 
These results do not show the presence of the cosmic ray shadows on their own (and would require significant more time to do so) but they have been used to verify that the tracking and the analysis method is working correctly.

\section{Lunar Eclipse Observations}
\begin{figure}[h]
\centering
\includegraphics[width=0.7\linewidth]{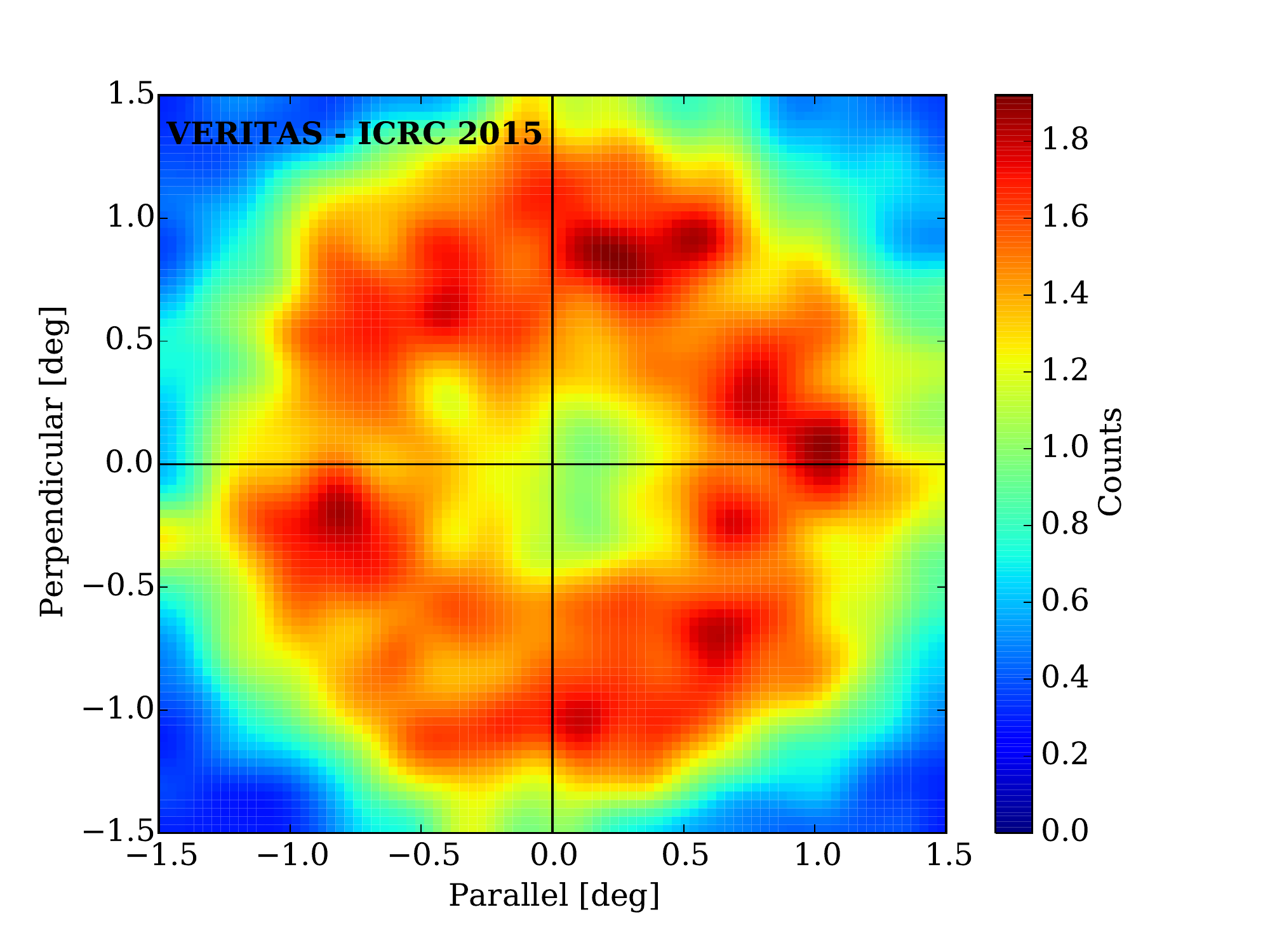}
\caption{The 20725 showers that passed the quality cuts are plotted relative to the Moon in the coordinate system described in the text with a \ang{0.1} Gaussian smoothing applied.  The deficit due to the Moon is slightly offset from the center of the camera.}
\label{Fig:LunEclMap}
\end{figure}

On 2014-4-15 a total lunar eclipse was visible from the VERITAS facility.
74 minutes of observations were taken pointing directly at the Moon  in the elevation range 45 to \ang{50}.
During the majority of the time the currents were low (less than background stars that do not normally cause an effect on the analysis).
The last 14 minutes had slightly higher currents, reaching levels comparable with bright stars towards the end of the data taking and thus this time has not been used in this analysis.
Quality cuts were also chosen to minimise the impact of any optical brightness (at the cost of statistics and a raised energy threshold) by using a large \textit{Size} (number of digital counts in the cleaned image) cut and requiring that a minimum of three telescopes are used in the image reconstruction.
Up to the event reconstruction stage the standard analysis packages are used, after that custom methods are employed (with an independent cross check) to plot the positions of the showers in a coordinate system centred on the Moon, but with the axes defined by the midline of the shadow (\textit{Parallel}) and its perpendicular (\textit{Perpendicular}).
The \textit{Parallel} distance is then normalised so that the distance between the \SI{1}{TV} shadow position and the Moon is constant.
The results are shown in Figure \ref{Fig:LunEclMap} where a deficit due to the shadowing of the Moon is clearly visible slightly offset from the Moon along the \textit{Parallel} axis.

To test for deflection, a fit was conducted using a 2D Gaussian over the central part of the FoV.
This fit gives a centroid location of (\textit{Parallel, Perpendicular}) = (\ang{0.09}, -\ang{0.06}), $\sigma$ = (\ang{0.19}, \ang{0.20}).
It should be noted that the energy threshold is estimated at above \SI{4}{TeV} at which energy the shift is expected to be \ang{0.2}.
Further work is ongoing to determine whether there is any evidence for deflection. 
If selection cuts are applied to examine the electron shadow then only 32 counts remain and no conclusions can be drawn. 

\section{Conclusion}
VERITAS is using the EMIS technique to measure the positron fraction in the energy range \SIrange{0.5}{1}{TeV}; a measurement which has been of great interest to the community over the last few years.
This measurement will require many tens of hours, (if not hundreds for the positron shadow) taking many years to accumulate the data.
Observations conducted during a lunar eclipse and pointed directly at the Moon show a clear deficit in the cosmic ray flux due to the shadowing of the Moon.
Work is ongoing to determine whether there is any evidence for a deflection in the shadow position from the Moon.

\acknowledgments
This research is supported by grants from the U.S. Department of Energy Office of Science, the U.S. National Science Foundation and the Smithsonian Institution, and by NSERC in Canada. 
We acknowledge the excellent work of the technical support staff at the Fred Lawrence Whipple Observatory and at the collaborating institutions in the construction and operation of the instrument.
R. Bird is funded by the DGPP which is funded under the Programme for Research in Third-Level Institutions and co-funded under the European Regional Development Fund (ERDF).
The VERITAS Collaboration is grateful to Trevor Weekes for his seminal contributions and leadership in the field of VHE gamma-ray astrophysics, which made this study possible.

\end{document}